# DETECTION OF INSIDER ATTACKS IN BLOCK CHAIN NETWORK USING THE TRUSTED TWO WAY INTRUSION DETECTION SYSTEM

D.NancyKirupanithi[1], Dr.A.Antonidoss[2] ,G.Subathra[3]

[1]Research Scholar & Computer Science and engineering &
Hindustan Institute of Technology and Science, Padur,

Chennai-603103,Tamil Nadu, India.

[2]Associate Professor & Computer Science and engineering &
Hindustan Institute of Technology and Science, Padur,

Chennai-603103,Tamil Nadu, India.

[3]Research Scholar & Computer Science and engineering &
Hindustan Institute of Technology and Science, Padur,

Chennai-603103,Tamil Nadu, India.

*Abstract—*

*For data privacy, system reliability, and security, Blockchain technologies have become more popular in recent years. Despite its usefulness, the blockchain is vulnerable to cyber assaults; for example, in January 2019 a 51% attack on Ethereum Classic successfully exposed flaws in the platform's security. From a statistical point of view, attacks represent a highly unusual occurrence that deviates significantly from the norm. Blockchain attack detection may benefit from Deep Learning, a field of study whose aim is to discover insights, patterns, and anomalies within massive data repositories. In this work, we define an trusted two way intrusion detection system based on a Hierarchical weighed fuzzy algorithm and self-organized stacked network (SOSN) deep learning model, that is trained exploiting aggregate information extracted by monitoring blockchain activities. Here initially the smart contract handles the node authentication. The purpose of authenticating the node is to ensure that only specific nodes can submit and retrieve the information. We implement Hierarchical weighed fuzzy algorithm to evaluate the trustability of the transaction nodes. Then the transaction verification step ensures that all malicious transactions or activities on the submitted transaction by self-organized stacked network deep learning model. The whole experimentation was carried out under matlab environment. Extensive experimental results confirm that our suggested detection method has better performance over important indicators (such as Precision, Recall, F-Score, overhead )*

## I. INTRODUCTION

In recent years, blockchain technology has seen explosive growth in the data security sector as a means of protecting sensitive information from being stolen or tampered with. Since its introduction in 2008, it has been used in a variety of settings with the primary goal of ensuring the reliability and consistency of data. The immutability of blockchain data and its distributed ledger technology make it ideal for data security. Since its creation, it has been used in several contexts, including but not limited to the medical field, data integrity protection, intrusion prevention, cooperative intrusion detection, and many more. Blockchain is a distributed, append-only public ledger that records network transactions in chronological order. Public and private Blockchains are the two main categories into which Blockchain may be categorized. All nodes in a public blockchain verify and validate transactions, making it a permission less distributed ledger. such as Bitcoin and Ethereum. Private block chains, on the other hand, are permissioned blockchains that only let selected nodes to take part in the network. such as Hyperledger. The term "nodes" is used to describe each and every member of a blockchain network. The blocks of information on a blockchain represent individual transactions. Continuity from one block to the next is essential (parent block). A block always includes a hash of the previous block as well as some information about the block. In this way, each block may reference its parent via a pointer. The majority of the system's users agree on the validity of each and every transaction in the public ledger. When a transaction block is added to a blockchain, it becomes hard to alter or delete the associated data. Blockchain technology successfully protected stored data against intrusions, but it has significant limitations, including the inability to detect and remove a hacked blockchain node. Research on blockchain's potential uses in the field of cyber security has mostly centered around protecting the information it stores from hackers. However, there has been minimal work done to develop methods for identifying internal assaults on databases. a network of verified blockchain nodes. Identifying and isolating rogue nodes may be difficult, particularly in a private blockchain network, despite the solution's potential benefits in protecting the reliability of data that is transmitted among participants. Continuous monitoring of the node's activity and the implementation of a server-side node authentication are required to identify a malicious insider node in a blockchain network. In addition to keeping an eye on how the nodes are behaving, every node's reported transactions should be checked for accuracy. Transactions in a Blockchain network are created, verified, and validated by nodes. These nodes, if hacked, may alter submitted transactions, insert phony transactions, or obtain illegal information, all of which can jeopardize the reliability of the stored data.

In this research, we present a unique architecture for continually monitoring the activity of each node in a blockchain to detect and prevent harmful acts by a compromised insider node. By monitoring historical attack data, the suggested architecture may identify hostile insider activity. There are three types of insider attacks that we often carry out: (1) Attacks designed specifically to disable the Blockchain. (2) Attacks that try to introduce spoofed information into the database, and (3) Attacks that try to steal or obtain sensitive information without permission.

The contributions of our work can be summarized as follows:

 • In order to identify rogue nodes in a blockchain network, it is necessary to offer a blockchain-based architecture capable of continually monitoring node activity.
 • By using a self-organized stacking network, the suggested architecture is able to identify nodes on the inside that are trying to detect assaults on the blockchain.
 • Using the Hierarchical weighed fuzzy algorithm, the framework can identify insider nodes attempting to insert false data into the database or obtain secret information.
 • The proposed system can confirm the authenticity of the supplied transaction (attack features/signature) and display it in a standard manner, which welcomes the involvement of heterogeneous IDS nodes.
 • The architecture can instantly and permanently communicate the confirmed transaction amongst IDS nodes through a distributed public ledger.

"This paper's remainder is organized as follows: Section II discusses the background and related works on blockchain application as IDS. Section III describes the problem statement, Section IV illustrates the proposed architecture. Section V presents the results, and finally in section VI presents the conclusions of this paper and possible future works".

## II. RELATED WORKS

When it comes to network connection, as is the case with contemporary information and block chain systems, the increasing automation and dependence on technology also increases the susceptibility and exposure to risk. Threats to information security may come from a wide variety of internal and external actors for a wide variety of reasons, including sabotage, data theft and destruction, fraud, hobbies, spying, state-sponsored crime and terrorism, political or military aims, and espionage. An insider is a human entity that has/had access to the information system of an organization and does not comply with the security policy of the organization, according to one definition, though there is no universally accepted definition of insider/insider threat in the context of information security. An employee, contractor, or other business partner who has or had authorized access to an organization's network, system, or data and intentionally exceeded or misused that access in a manner that negatively impacted the confidentiality, integrity, or availability of the organization's information or information systems is considered a malicious insider. The insider threat issue and the methods to address it have received increased attention over time. Numerous strategies have been presented in the efforts done to address the insider threat issue, with a focus on identifying the behavior of possibly harmful insiders.

In [1], the authors present an Intrusion Detection System that relies on a Collaborative Blockchain-based Signature (CBSigIDS). CBSigIDS is an open-source, blockchain-based, collaborative signature-IDS platform. This framework uses blockchains to gradually upgrade a trusted signature database for various nodes in an IDS network. CBSigIDS's effectiveness in preventing simulated assaults on collaborative intrusion detection systems or networks from threats like worms and floods was tested (CIDN). The assessment contrasted the outcomes of CIDN simulations with those of actual CIDN. The findings demonstrated that by creating a trustworthy signature database, blockchain technology may improve the resilience and efficacy of signature-based IDSs in hostile circumstances. The authors of [2] proposed a SectNet, an architecture for protecting data while it is being stored, processed, and shared in a wide-area network like the Internet. The design was created to make cyberspace safer by using real big data and improved AI with plenty of data. Trusted-value exchange platform, artificial intelligence-based secure computing platform, and blockchain-based data sharing with ownership guarantee are all components of their design. In order to gauge how well Blockchain performs, we looked at how susceptible it is to common network assaults like Distributed Denial-of-Service (DDoS) Attacks and how much money it generates for the people who help keep the network secure. The outcome shown that the SectNet greatly mitigated the effects of DDoS attacks by having all internet users share their security rule settings. After the quality impact of the real market has created, the contributor's income will rise at a greater pace if the shared security regulations are of better quality. The study reveals that this work is tailored to distributed denial of service assaults. The authors also did not discuss the validity checks performed on data stores. A blockchain anomaly detection system (BAD) designed to identify assaults on the blockchain network was presented in [3]. Because of BAD, a Blockchain cannot be further corrupted by the introduction of a rogue transaction. Forks are a kind of blockchain information that BAD uses to track possible harmful actions taking place in the blockchain network. Using machine learning, these projects taught blockchain nodes to spot suspicious behavior. The authors took into account the possibility of an eclipse attack (in which a malicious actor compromises a node's IP address list, hence gaining control over all IP addresses on the target node). Data analysis revealed that BAD could identify and halt attacks that use bitcoin splits to propagate malicious malware. However, this only addresses blockchain network assaults that use Bitcoin splits. Collaborative Internet of Things anomaly detection via a distributed ledger system is the topic of another study, suggested in [4]. (CIoTA). By using the blockchain, CIoTA is able to conduct anomaly detection on IoT devices in a decentralized and cooperative manner. Anomaly detection models were continually trained independently using CIoTA, and then their collective knowledge was utilized to distinguish between very uncommon, harmless occurrences and malicious ones. The results evaluated shown that coupled models could readily identify malware activity with no false positives. The suggested solution relies on a network of IoT devices working together to spot assaults, but it doesn't explain how to spot a device that's been tampered with. In [5], the authors suggest using blockchain technology to identify malware on mobile devices. They compiled a feature set by extracting data on Android malware families from installation packages, permission packages, and call graph packages. They found evidence that the approach could identify and categorize prevalent forms of malware. In addition, it can determine whether or not a piece of software is malicious and classify it into one of many different malware families more quickly and efficiently than before. The preceding fix is tailored to counter host-based malware attacks on Android-based mobile devices. Therefore, extending it to network-based assaults will be challenging. Despite several studies, existing systems prioritize security for the shared data. Conversely, spotting and isolating compromised insider risks receives less focus. Our suggested architecture is able to identify compromised insider nodes and then isolate them via constant monitoring of the nodes' activities. However, the method also finds external dangers. The aforementioned justification both inspired this study and sets it apart from others in the field. The author of [6] created a consortium blockchain network to compare different machine learning models against a standard malware dataset. Smart contracts incentivize participants for their efforts by providing a safe and reliable means for them to submit solutions gained via training with predetermined ML models. Competitors' examination of shared information benefits all firms in the network by providing insights into how to improve existing malware detection and security mechanisms. The decentralized network eliminates the need for intermediaries, which increases openness, security, and efficiency when maintaining all important data. Results from preliminary experiments using the created framework and the DREBIN dataset are provided, and they are found to be promising. In [7], the writer suggests a blockchain-based firmware upgrading platform to improve the current system. Integrity checks and virus screening are made mandatory with the use of a smart contract. Our platform is less vulnerable to DDoS assaults because of the peer-to-peer file sharing technology we've implemented. In the event of several updating requests, we use batch verification for improved scalability. In [8], the authors suggest two innovative ideas: I a user (local) neural network (LNN) that trains on local distribution, and (ii) a smart contract that enables an aggregation process via a blockchain platform, with the latter ensuring the model's validity and quality. While the LNN model investigates a wide range of static and dynamic characteristics shared by malicious and benign programs, the smart contract checks for the presence of malware during the uploading and downloading phases of network operations by drawing on a cache of aggregated features from local models. So, the suggested approach not only enhances model effectiveness using blockchain, but also increases malware detection accuracy through a decentralized model network. They conduct in-depth assessments of the retrieved characteristics of the each model and compare the results to those of three state-of-the-art models, all of which are used to evaluate their technique. To stop malware insertion into a blockchain network, the authors of [9] suggest a concept that uses a convolutional neural network in conjunction with a blockchain. Prior to adding data to the blockchain, this convolutional neural network checks for the presence of malicious code. They examined two distinct CNN models, one based on the VGG-16 architecture and the other based on a more streamlined, individualized approach with fewer layers. In [10], we investigate post-authentication vulnerabilities and provide a blockchain-based, reputation score-based architecture. A device's reputation is calculated from the feedback of its neighbors. In order to guarantee the honesty, accessibility, and decentralization of the views and incentives, they are recorded on a blockchain. By extending the publicly available Go implementation of Ethereum, they test the proposed model for latency, and then evaluate the model's security and efficacy using a reputation simulator. The authors of [11] described a blockchain-based database that encrypts and stores information. Information on the blockchain network is stored and shared using an interface, the Web Application Programming Interface (API). The adversary may use a variety of assaults on the database to weaken the defenses of the system. In [11], the authors develop a blockchain-based filtration mechanism with CIDN to aid in protecting the security of IoT networks by refining unexpected events; in [12], they focus on the SQL injection attack that the adversary performs on Web API and present a case study based on the Snort and Moloch for automated detection of SQL attack, network analysis, and testing of the system. To host and distribute data like blacklist, we use IPFS technology as well. The authors of [13] present an edge-based blockchain-enabled anomaly detection approach to avoid insider assaults in the Internet of Things, and they conduct an assessment of the filter's effectiveness using three genuine datasets, a simulated environment, and a practical environment. Edge computing is used initially to bring processing closer to the IoT nodes, which decreases latency and bandwidth needs, increases availability, and eliminates single points of failure. Finally, it incorporates a distributed edge with blockchain, which provides smart contracts to execute anomaly detection and correction in incoming sensor data, capitalizing on some part of sequence-based anomaly detection in the process. In [14], the author made an effort to unearth insider threats by tracking down out-of-the-ordinary actions taken by company personnel throughout their social media and other digital channels. The team analyzed and gleaned elements that are predictive of insider threat activity and processed them accordingly. However, the major emphasis of their efforts was on insider threats that could be linked to specific workers by their digital footprints. It didn't take into consideration circumstances where no digital footprints are left, nor did it take into account the IoT domain, which is a highly automated system with no human participation and deals with real-time data. Using log analysis and event correlation, [15] suggested a strategy for detecting insider threats. They proposed a statistical approach to prove the frequency of recurrence of an event by looking at past records of that event. The saved log file served as the primary input to the event correlation system, which then calculated the likelihood of hostile insider activity as a percentage. Furthermore, the Internet of Things (IoT) is not a good fit for this method since it needs a way to verify the validity of incoming data before acting on it. Both [16] and [17] provided schemes outlining models and frameworks to better comprehend insider dangers, citing other relevant publications in the process. Additionally, studies like [18] focused on foreseeing and avoiding the insider threat issue. Some solutions were suggested in the examined literature, but they did not take into account insider threat issues in the IoT sector. Contract Guard was designed by the author in[19] and implemented in contracts to profile context-tagged acyclic pathways, with optimizations made according to the Ethereum gas-oriented performance model. With digital concurrency, the initial cost is a major consideration, therefore reducing the associated overheads is a top priority. In [20], the author gathers real-world examples and suggests a method for identifying blockchain-based Ponzi schemes carried out by use of smart contracts. Before anything else, over 3,000 open source smart contracts on the Ethereum platform are carefully checked, yielding a total of 200 intelligent Ponzi schemes.

There are thus two types of characteristics retrieved from the smart contracts' operation codes and transaction histories. To conclude, a classification model is provided for identifying sophisticated Ponzi schemes. In [21], the author offers a unique approach to identifying hacked blockchain nodes by preventing their actions from being updated in the blockchain ledger using a server-side authentication mechanism. They evaluate the proposed system by carrying out four typical insider assaults, which may be sorted into the following three groups: (1) Attacks aimed directly against the Blockchain, which would cause it to crash. (2) Attacks that try to corrupt a database by inserting bogus information. Thirdly, intrusions meant to steal or obtain private information. They explained the technical details of the assaults and how our design can identify them before they have an effect on the network. Finally, they compared our system to others already in use and showed how long it took to identify each assault.

### III. PROBLEM STATEMENT

This effort was prompted by the ever-increasing security challenge to the block chain system, which ensues from the numerous uses of the system. Those widespread applications include: The most important factor, however, is the existence of Insider Threats, which refer to the fact that system insiders often hold advantageous positions inside the system. As a result, systems become more vulnerable to exploitation and assaults when insiders hold these positions. The emphasis placed on the identification of anomalies is warranted since the most effective method for efficiently securing any system is the early detection and avoidance of ongoing threats to the system or assaults on the system. There have been a number of other methods suggested, but they all have some of the same drawbacks.

### IV. PROPOSED WORK

We argue that using trust as a tool for assessing a node's behavior across several dimensions is an effective strategy for countering sophisticated insider assaults. Through this endeavor, we want to establish a block chain-based trust by creating a chain of confirmed hostile comments. When a node engaged in testing gets input, it may then pass that information along to other nodes. Then, every node must decide whether or not to upload the new information to the block chain based on whether or not there is an anomaly in the shared pair. In order to determine whether a response is malicious, we look for a large discrepancy between the response to the challenge and the response to the request. The pictorial architecture of the suggested system was depicted in figure 1.

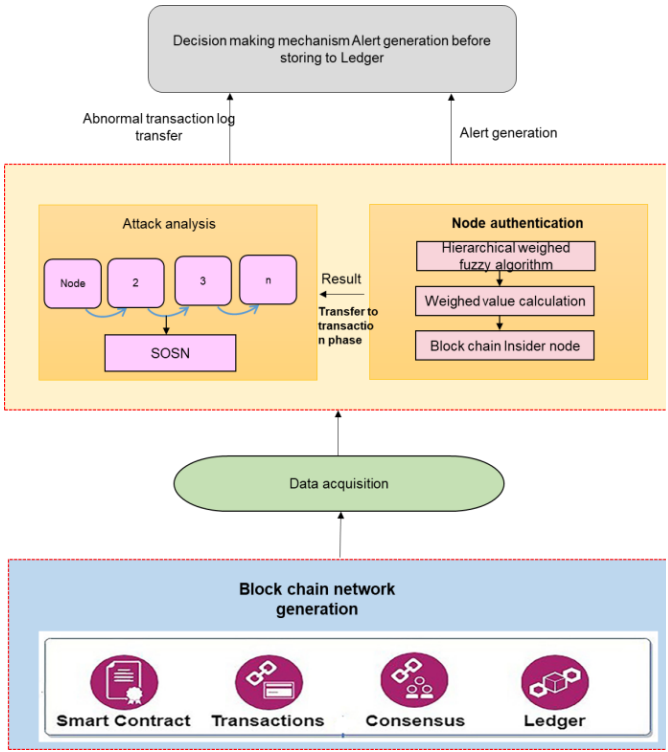

**Figure 1 schematic representation of the suggested methodology**

### a. Data acquisition

To express the network topology, we use the formula $\mathcal{H} = (\mathcal{U}, \mathcal{F})$, where $\mathcal{U} = \{\mathcal{U}_1, \ldots \mathcal{U}_M\}$ is a collection of all nodes with $|U| = M$, and F is an arbitrary set of vertices. The communication path in the network is represented by the set of edges $\mathcal{F} \subseteq \mathcal{U} \times \mathcal{U}$ where $(\mathcal{U}_i, \mathcal{U}_j) \in \mathcal{F}$. If there is a communication connection $\mathcal{U}_i, \mathcal{U}_j)$ between U i and U j, then U $_i$ and U $_j$ may have a two-way conversation with each other. A node i's neighbour set is defined as

$$\mathcal{M}_i = \{\mathcal{U}_j \in \mathcal{U} : (\mathcal{U}_i, \mathcal{U}_j) \in \mathcal{F}\}, \forall j \in \{1,2,\ldots,M\} \quad (1)$$

By default, the network's nodes all engage in a kind of random chatter communication. For a node I to update its local state information, it must communicate with its immediate neighbour $j \in \mathcal{M}_i$. We suppose, without sacrificing generality, that this consensus procedure is executed in a total of L separate instances. Let $y_i^l(0) = \gamma_i^l$ be the starting condition for a node I that behaves normally, where $\gamma_i^l$ is a discrete random variable that remains constant throughout time. For a typical consensus procedure, the recursion,

$$y^l(s+1) = P^l(s)y^l(s), \quad s = 1,2,\ldots,S. \quad (2)$$

$P^l(s) \in \mathbb{S}^{M \times M}$ is the weight matrix at instance l and time s, and $y^l(s) = (y_1^l(s), \ldots, y_M^l(s))^S \in \mathbb{S}^M$ is the random vector representing the states of all the nodes at the sth iteration. From now on, we will no longer write P(s) with a superscript l, since that prefix is unnecessary. A node I chooses a node j at random, and their states are synchronised such that they are identical to the mean in the sth time frame. Here is the weight matrix:

$$P_{ij}(s) = I - \frac{(f_i - f_j)(f_i - f_j)^S}{2} \quad (3)$$

"where $f_i = [0, \cdots, 0, 1, 0, \cdots, 0]^S$ is an M × 1 unit vector with the ith element equal to 1. By defining $[A]_{ij} = A_{ij}$ and $\Sigma = diag([\Sigma_1, \ldots, \Sigma_M])$ as a diagonal matrix with $\Sigma_j = \Sigma_{j=1}^M (A_{ij} + A_{ji})$, the expected weight matrix is":

$$\bar{P} = \mathbb{F}[P(s)] = I - \frac{1}{2M}\Sigma + \frac{A + A^S}{2M} \quad (4)$$

"This is demonstrable $\bar{P}$ is asymmetrical and stochastic in two ways without being negative. Predicted values may be written out as,

$$\mathbb{F}[y^l(s)|y^l(0)] = \bar{P}\mathbb{F}[y^l(s-1)|y^l(0)] = \bar{P}^s y^s(0) \quad (5)$$

"Suppose the network is connected in expectation, i.e., if $\lambda_2(\bar{P}) < 1$. The state of node i $\in \mathcal{U}$ converges to the average of network initial states"

$$\lim_{s \to \infty} y^l(s) = 1^S y^l(0)/M. \quad (6)$$

### b. Node authentication

We focus on the authentication level, trust level, traffic volume, and node activity level of each node as the primary criteria for identifying intrusions with a high degree of accuracy. The suggested scheme's primary objective is to guarantee that a prospective new customer poses no threat to the cloud infrastructure when they wish to enter the market. The system functions on two tiers: the cloud and the client. The new node has to pass the proposed two stages of authentication before it can enter the cloud market. Each client's certificate is checked for validity in the cloud. Each customer in the market receives a unique certificate, issued by the cloud. Each certificate has its own public and private keys, as well as a unique logical identification and media access control (MAC) address. Since the cloud service provider (CSP) issues certificates for each node, each client in the market must contact the CSP to verify certificates for other nodes. Except for the node's private key, which is never divulged to any other node in the market, the CSP must provide each client all of the information they need about the node in order to authenticate it. Before letting a new node into the cloud market, CSPs should do virus and hacking tool scans to prevent malicious assaults. The new node must provide the certificate issued by its CSP to the CSP, which will only work with CSPs that are on a specified list. A fresh certificate is issued to the node if it was issued by a trusted certificate service provider (CSP). Next, the client transmits the new certificate in an encrypted message together with the node ID and MAC address using the public key. For the purposes of a Hierarchical weighed fuzzy algorithm, the following is the definition of authentication's evaluation value,

$$P_s = \begin{cases} 1, & authenticate\ node \\ 0, & o.z \end{cases} \quad (7)$$

"Fuzzy measure $S_x$ is defined on P(Y) of a finite set Y satisfying the following properties:
- $S_x : P(Y) \to [0, 1]$
- $S_x(\emptyset) = 0$, $H_x(Y) = 1$
- If $P, Q \in P(Y)$, $P \subset Q$, then $S_x(P) \leq H_x(Q)$.
- If $G_M \in P(Y)$, $\forall 1 \leq m < \infty$ and a sequence $\{G_M\}$ is monotone where $\lim_{m \to \infty} H_x(G_M) = H_x(\lim_{m \to \infty} H_x)$"

There are several factors to take into account while determining the confidence in each node using fuzzy properties. Some examples of such metrics include reaction time, throughput, availability, and success rate. The CSP maintains a record of these properties and their values. These settings are refreshed after every market occurrence, so you can be certain that you're always using the most up-to-date parameters possible. The response time for a node is the total amount of time it takes to process a service request. Here's how you may calculate the i-th node's response time:

$$S_j = \frac{S_r - S_{min}}{S_{max} - S_{min}} \qquad (8)$$

where $S_{max}$ and $S_{min}$ represent the highest and lowest response times in the area, and $S_r$ represents the average response time of a node. The more suspicious a node seems to be, the higher its $S_i$ score. A node's throughput is the proportion of its transmissions that are received successfully. For the i-th node, throughput is determined by the formula:

$$G_j = \frac{G_r - G_{min}}{G_{max} - G_{min}} \qquad (9)$$

where $G_{max}$ and $G_{min}$ represent the highest and lowest achievable throughputs in the area, and $G_r$ represents the throughput of an average node. The lower the node's $G_i$ value, the more probable it is to be an invader. "Availability" refers to how long a system is fully operational when it is needed to be. To keep the node operational, the CSP would work to fix the problem. However, frequent failure of a node suggests that it is behaving deliberately. As a result, the following formula may be used to determine availability's valuation in an evaluation:

$$U_j = 1 - \frac{U_0 - U_{min}}{U_{max} - U_{min}} \qquad (10)$$

where $U_0$ represents the sum of the i'th node's failure occurrences, $U_{max}$ represents the most number of restarts received from neighbors, and $U_{min}$ represents the smallest. Success rate refers to the proportion of requests that are actually satisfied. Consequently, the success rate's evaluation value may be calculated as follows:

$$T_j = 1 - \frac{T_0 - T_{min}}{T_{max} - T_{min}} \qquad (11)$$

where $T_0$ represents the total number of requests processed by the i-th node, $T_{max}$ the largest number of requests processed from neighbors, and $T_{min}$ the smallest. High contention rates occur when customers are unable to access cloud services due to an invader inner node in the cloud market pumping more traffic into the network than it can manage. After numerous failed attempts to relay a packet by one of the later nodes, the connection is marked broken, and the routing algorithm begins searching for a new route. Until a different path is found, cloud market packets cannot be transmitted. Because of this, packet loss rises and throughput drops significantly. As more superfluous traffic is pumped into the system, network latency increases and service quality declines. Each node was assigned a trust score based on the findings of the research.

$$P_s = \begin{cases} 1, & j \in O \\ 0, & o.z \end{cases} \qquad (12)$$

where $P_s$ is the nodal trust score, O denotes the set of trust distances.

Node ranking can be computed as follows:
$$K_j = \Sigma_{S \to \infty} P_s \qquad (13)$$

where $K_j$ denotes the node ranked matrix

This is how we may anticipate the suspicious actions of a node:

$$M_j = \frac{K_j - K_{min}}{K_{max} - K_{min}} \qquad (14)$$

In this equation, $M_j$ represents the i-th node's activity level, $K_{max}$ and $K_{min}$ represent the maximum and lowest numbers of activities in which neighbors take part, and n represents the set of all possible activities.

c. Attack analysis

Our concept characterizes adversarial nodes as those whose states are immutable by other nodes. Separating the nodes into a trusted set $U_r$ and an untrusted set $M_j$ allows us to identify the insider node and anticipate the assault.

i.e., $\mathcal{U} = \mathcal{U}_r \cup M_j$, and $M = |\mathcal{U}_r| + |M_j|$. Any malicious updates from node j are handled according to a stacked network rule.:
$$y_j^l(s + 1) = \alpha^l + n_j^l(s). \qquad (15)$$

where $\alpha^l$ is the attackers' targeted value at the lth occasion and $n_j^l(s)$ is traffic created by the attackers to throw off the network's detection mechanisms. The attackers' activity is masked as that of ordinary nodes by the exponential decay of the fake noise, which mimics the converging tendencies anticipated by the algorithm. Let's break out the whole state vector like so to define the attack model:

$$y^l(s) = (t^l(s)^S r^l(s)^S)^S \qquad (16)$$

where $t^l(s) \in \mathbb{S}^{|\mathcal{U}_s|}$ and $r^l(s) \in \mathbb{S}^{|\mathcal{U}_r|}$ are the attacker and normal node state vectors, respectively. That no other attackers may alter a given attacker's state is represented by the weight matrix. $P_{ij}(s) = 1 \; if \; i = j \epsilon \mathcal{U}_s$, and $P_{ij}(s) = 0 \; if \; i \epsilon j, \mathcal{U}$. As a result, the block structure is allowed in the predicted weight matrix $\bar{P}$.

$$\bar{P} = \begin{pmatrix} I & 0 \\ \bar{Q} & \bar{C} \end{pmatrix} \qquad (17)$$

where $\bar{Q} \in \mathbb{S}^{|\mathcal{U}_r| \times |\mathcal{U}_s|}$, $\bar{C} \in \mathbb{S}^{|\mathcal{U}_r| \times |\mathcal{U}_r|}$ are the submatrices between the attacker and regular nodes, and between the normal

nodes themselves. If they are successful in this situation and remain undiscovered, the attackers will have influenced the outcome.

To achieve the target value, the state of node i ∈ $\mathcal{U}$ in all l instances converges under the attack model $\alpha^l$, i.e.,
$$\lim_{s \to \infty} y^l(s) = \alpha^l 1 \quad (18)$$
This research considers the detection and localization challenges from the perspective of a self-organized stacking network. The first order of business is to check whether any attackers are present in the network by examining the stacked layer 1 nodal trust value. For the most part, each regular node in the network performs the duty of network detection independently i ∈ $\mathcal{U}_r$, i.e.,

"$G_0: \mathcal{U}_s = \emptyset$, There is no attacker in the network,
$G_1: \mathcal{U}_s \neq \emptyset$, Attacker is present in the network".

The second mission in $G_1$ is to check for hostile nodes near node i. The following is a definition of the layer 2 neighborhood detection problem:

"$G_0^i: \mathcal{M}_i \cap \mathcal{U}_s = \emptyset$, No neighbor is an attacker,
$G_1^i: \mathcal{M}_i \cap \mathcal{U}_s \neq \emptyset$, At least an attacker neighbor".

When $G_1^i$ occurs at node I the third mission is to identify the perpetrator among the nearby nodes. For the sake of this paper, we will define the neighborhood localization as

"$G_0^{ij}: j \notin \mathcal{U}_s$, Node j is not an attacker,
$G_1^{ij}: j \in \mathcal{U}_s$, Node j is an attacker., ∀ j ∈ $\mathcal{M}$"$_i$.

If $G_1^{ij}$ If correct, then the attacker is restricted to a certain area. As a result, node I will sever any further communications with the adversary.

We refer to $y_i^l$ as the trust value in the state vector that node I has accumulated in the lth instance, where $y_i^l$ is the state of node j ∈ $\mathcal{M}_i$, which node I may directly get. Below is a breakdown of the detection and localization processes:

$$y_i^l := [y_j^l, y_1^l, \ldots, y_j^l, \ldots, y_{|\mathcal{M}_i|}^l]^S \quad \forall j \in \mathcal{M}_i \quad (19)$$

$$c^i = CS(y_i^1, \ldots, y_i^l) \overset{G_1^i}{\underset{G_0^i}{\gtrless}} \delta, \quad \kappa_j^i = KS(y_i^1, \ldots, y_i^l) \overset{G_1^{ij}}{\underset{G_0^{ij}}{\gtrless}} \epsilon \quad (20)$$

"Where $k^i = [k_1^i, \ldots, k_{|\mathcal{M}_i|}^i]^S$ for localisation purposes, is the metric vector of nearby neighbors.

Let's pretend we have some background knowledge about the average attacked value and the average starting states of the normal nodes. "Usually, $\mathbb{F}[y_t^l(0)] = \bar{\alpha} \neq \bar{\gamma} =, \mathbb{F}[y_j^l(0)]$ the expected initial value of an attacker $t \in \mathcal{U}_s$ is different from the normal node $\in \mathcal{U}_r$. When $t \to \infty$ the network converges to $\mathbb{F}[y_t^l(\infty)]$. Then each normal node $\in \mathcal{U}_r$, in a decentralized fashion, evaluates the following $score$ ":

$$\xi_{ij} := \frac{1}{L}\sum_{l=1}^{L}(y_j^l(S) - y_j^l(0)), \quad j \epsilon \mathcal{M}_i \quad (21)$$

Herein, $y_j^l(S), y_j^l(0)$ are the most recent and earliest values of state recorded for node j close to node i. Following are some of the detection criteria that may be utilized to determine whether there is an active attacker on the network:

$$C_1^i := \frac{1}{|\mathcal{M}_i|}\sum_{j \epsilon \mathcal{M}_i} |\xi_{ij} - \bar{\xi}_i| \overset{G_0^i}{\underset{G_1^i}{\gtrless}} \delta_1 \quad (22)$$

"where $\bar{\xi}_i = (\frac{1}{|\mathcal{M}_i|})\sum_{j \epsilon \mathcal{M}_i} \xi_{ij}, \delta_1 > 0$ is a pre-designed threshold. The intuition underlying is that E[D i 1] = 0 when the attacker is not present while $\mathbb{F}[C_1^i] \neq 0$ otherwise".

For localization task, the $G_1^{ij}$ and $G_0^{ij}$ One way to distinguish one occurrence from another is to examine certain defining characteristics.

$$K_1^{ij} := |\xi_{ij}| \overset{G_1^{ij}}{\underset{G_0^{ij}}{\gtrless}} \epsilon_I, \forall j \epsilon \mathcal{M}_i \quad (23)$$

"where $\epsilon_I > 0$ is a pre-designed trust value".

Finally the insider attack can be predicted precisely. When the IDS identifies aberrant activity, it will undo all of the modifications that were made to the contract. states and sound the alarm to the administrators of the system. Because of this, an opportunity to avoid suffering an irreparable loss due to the vulnerability may cause.

---

**Algorithm 1: HWFA_SOSN**

**"Inputs: Ethernum network
Outputs: Insider attack prediction**
$\mathcal{H} = (\mathcal{U}, \mathcal{F})$ matrix of pairwise comparisons for nodal attributes.
w: matrix of wights for node attributes.
Tn
$\mathcal{H} = (\mathcal{U}, \mathcal{F})$: Trust value for nth node.
 P = Genrate- pair-wise()
    for i ← 0, N do
. t=1 (Time slot)
    for j ← 0, N do
$S_x: P(Y) \to [0, 1]$
• $S_x(\emptyset) = 0, H_x(Y) = 1$
• If P, Q ∈ P(Y), P ⊂ Q, then $S_x(P) \leq H_x(Q)$.
• If $G_M \in P(Y), \forall 1 \leq m < \infty$ and a sequence $\{G_M\}$ is monotone where $\lim_{m \to \infty} H_x(G_M) = H_x(\lim_{m \to \infty} H_x)$
    end for
$K_j = \Sigma_{S \to \infty} P_s$ (Rank)
End if
    Else
        Move on
    end for
 Sum = Score
    for i ← 0, N do
Attack mod$(y^l(s) = (t^l(s)^S r^l(s)^S)^S$          )
sum
[ ]
 end for
    for i ← 0, X do// Computing insider attack

end for
Accept=Node() else Reject-Node();

$$K_1^{ij} := |\xi_{ij}| \begin{smallmatrix} G_1^{ij} \\ > \\ < \\ G_0^{ij} \end{smallmatrix} \epsilon_I, \ \forall j \epsilon \mathcal{M}_i$$

End"

## V. PERFORMANCE ANALYSIS

We test the suggested security scheme to identify the attacker nodes that degrade network performance and cause insider attack. Figure 1 illustrates the network that was simulated under MATLAB environment, together with the parameters' values that were used. The significance and implications of using our scheme to protect data across block chain are highlighted through an evaluation of the results obtained through monitoring node activity and analyzing node data.

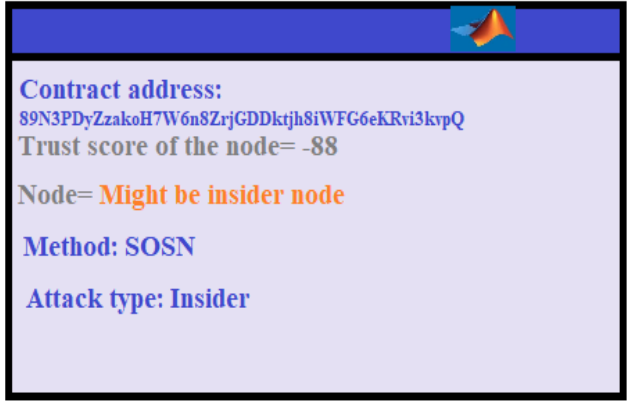

**Figure 2 Simulated output**

To prove the efficiency of the suggested algorithm it can be compared with the existing methods [19,20,21].

*1) "Deployment overhead:* After instrumentation, the contract code grows in size as the anomaly detection capabilities and the default safe route set are added. Since Ethereum imposes fees during contract generation, this will increase the overall cost of deployment."

*2) "Runtime overhead:* During transaction execution, the integrated IDS profiles the control flow pathways. Additionally, the IDS must validate path membership when a path dies".

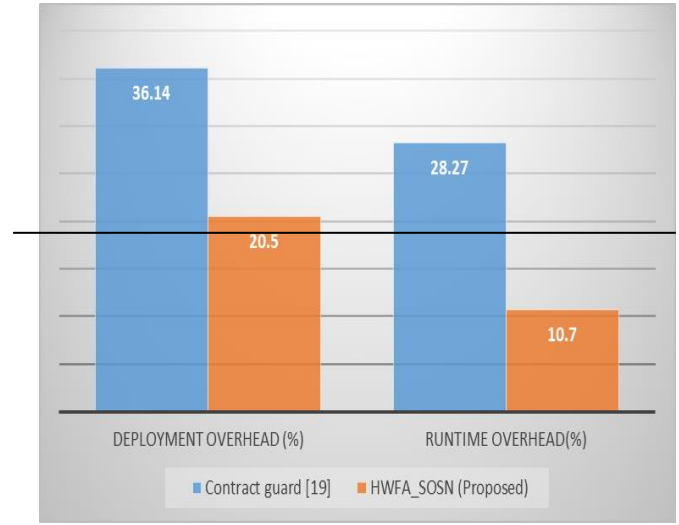

In Figure 3, we can see the relative size of the deployment and runtime overhead for the contracts. The average deployment and runtime overhead is 20.5% and 10.7% which is very low when compared to the existing mechanisms.

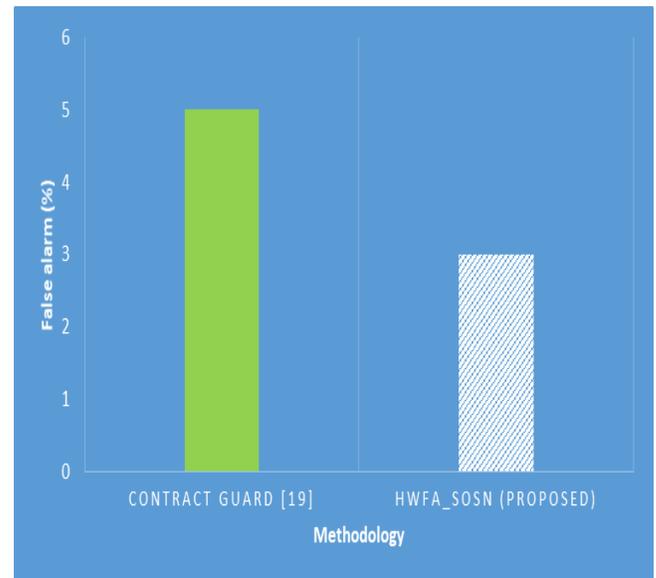

**Figure 4(a) False alarm rate**

False alarms, which represent legitimate transactions that have been incorrectly classified as malicious, are another important issue that might hamper the practical application of the proposed methodology. A false alarm can cause future legitimate transactions to be rejected unless the administrators manually review each alarm and add the path to the safe path set. There may be a lot of manual labor involved. The number of false alarms needs to be looked into. We assume that the developers deploy their contracts without training. Therefore, the technique will have an empty safe route created initially. We also presume that all transactions are valid, and that when a suspicious one is reported, the administrators will quickly add the transaction's context-tagged acyclic paths to the safe path set. Under these assumptions, we then count the amount of false alarms for each contract. This is a rough estimate of the maximum amount of human labor that would be needed to deal with false alarms in the real world.

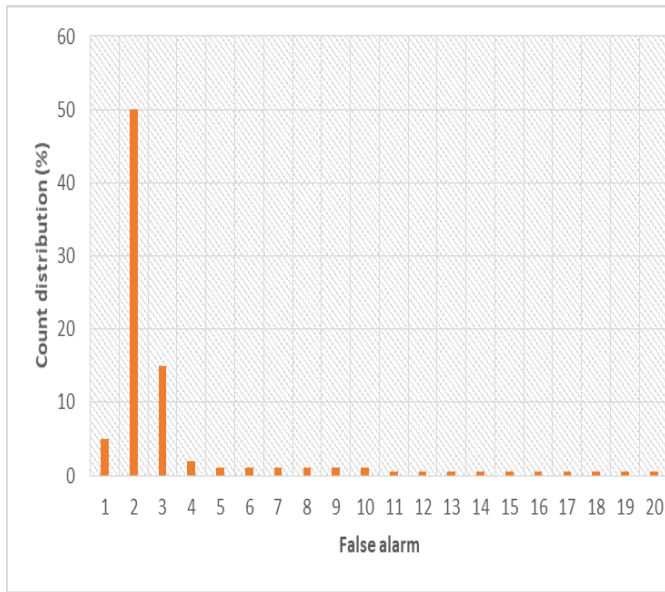

**Figure4(b) False alarm count distribution**

For a majority of the transactions the number of 3 but the existing method have a false alarm rate of 5.

To prove the efficiency of the classifier performance over attack detection it can be evaluated with some performance metrics below,

**Precision**

It determines how accurate the suggested technique's behavior is by identifying the insider attack from the block chain

$$\text{Precision} = \frac{TP}{TP+FP}$$

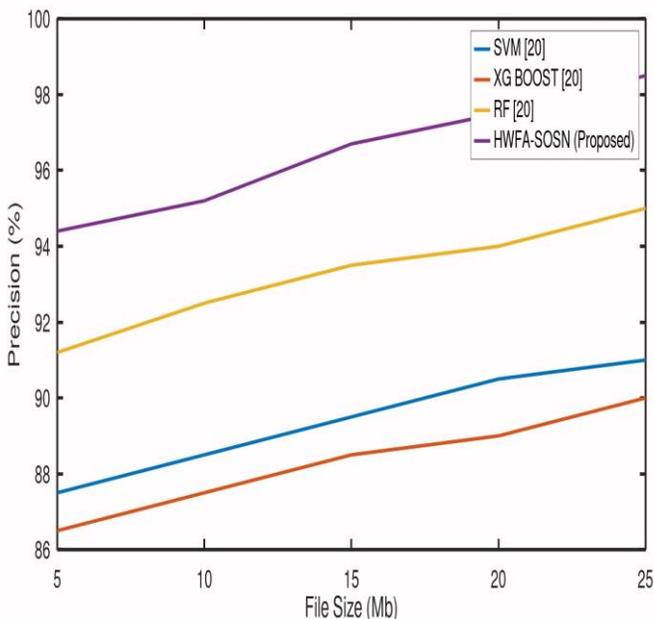

**Figure 5 File size Vs. Precision**

According on the results shown in Figure 5, the suggested HWFA)SOSN technique has a higher precision of up to 91.5 percent which is very high that that of the other existing mechanisms.

**F1 score**

With the help of two measures (precision and recall), we can easily calculate the F-measure. The F-score is calculated as,

$$F = \frac{2*Recall*Precision}{Recall+Precision}$$

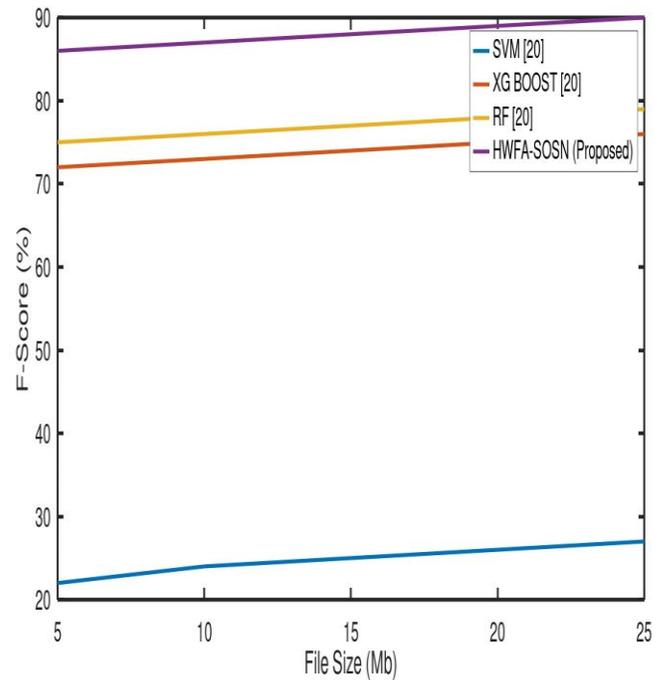

**Figure 6 File size Vs. Recall**

According on the results shown in Figure 6, the suggested HWFA)SOSN technique has a higher F1 score of up to 90 percent which is very high that that of the other existing mechanisms in use

**Recall**

The ratio of correctly predicted instances and all instances.

Recall= TP/(TP+FN)

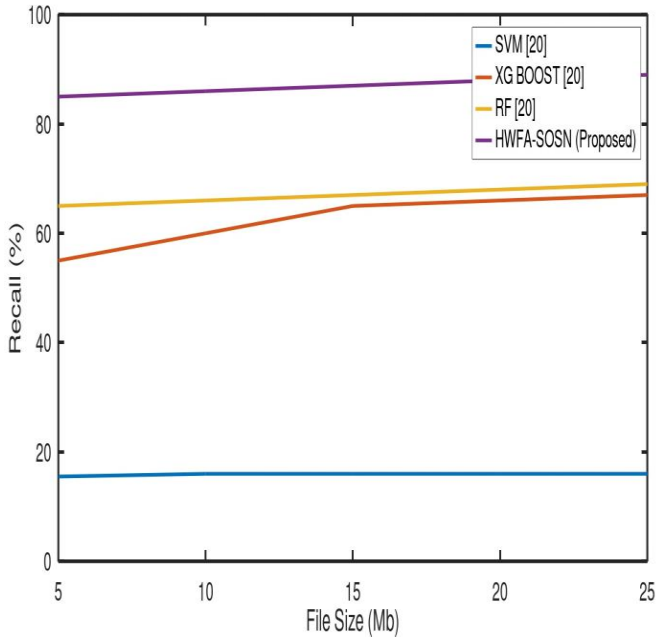

**Figure 7 File size Vs. Recall**

According on the results shown in Figure 5, the suggested HWFA)SOSN technique has a higher recall of up to 83.5 percent which is very high that that of the other existing mechanisms.

We analyzed how long it takes for attacks to be uncovered. Time to detection is the amount of time it takes for the systems to send out a warning about a failed transaction. It takes this long from the moment a transaction is filed to the time a notice is received. Each attack has a time stamp showing when it was submitted and when it was notified. The lag between the two is the detection time. The time it takes to detect the various attacks is depicted in Fig. 8. The time it takes to detect an insider attack that involves multiple transactions is the longest that we've seen. Reason for the longer period is that we waited to send the failure signal until after the smart contract counted the number of transactions per second. At last, we analyze how our method stacks up against the approaches taken by other related works (Table I). The result shows a clear distinction between our work and other related works.

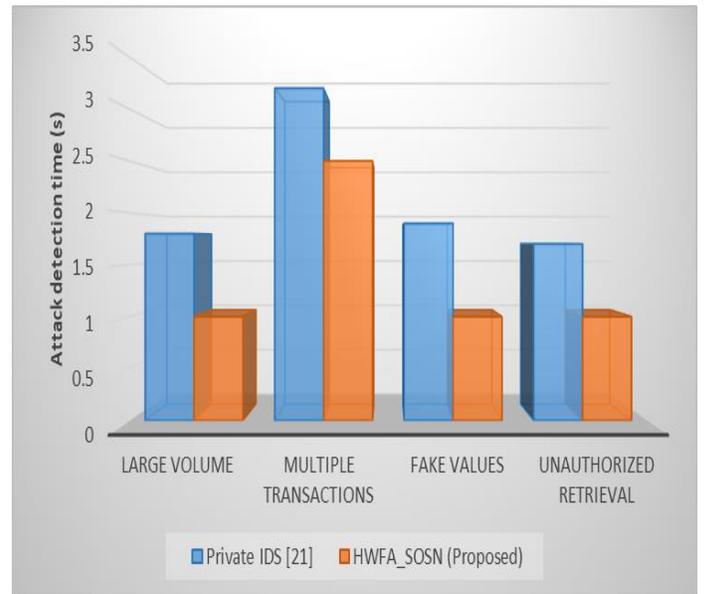

**Figure 8 Attack detection time**

As of from the figure 8 the time taken by the suggested classifier for block chain attack detection was an average of 2.5 second which is very low that that of the existing mechanisms.

**Table 1 Security analysis**

| "properties" | "Trust chain"[21] | "CBsigl DS"[21] | "Sectnet"[21] | Proposed [21] |
|---|---|---|---|---|
| Sharing data | ✓ | ✓ | ✓ | ✓ |
| BC | ✓ | ✓ | ✓ | ✓ |
| Insider node detection | x | ✓ | ✓ | ✓ |
| Security | x | ✓ | x | ✓ |
| SC verification | x | x | x | ✓ |

From the result obtained it was revealed that the suggested mechanism outperforms well over insider attack prediction that that of the existing mechanisms.

## VI. CONCLUSION

In this research, we have suggested two novel security strategies that are based on deep learning, and their goals are to identify and pinpoint insider nodes. The self-organized stacked networks are given their training at a centralized location, and then the models are distributed among the many network nodes for online detection. The experiments clearly show that the suggested Hierarchical weighed fuzzy algorithm along with the self organized stacked networks methods provide better performance in comparison to statistical detectors that rely on simplified expert models; in addition, they exhibit good robustness against model mismatch and continue to be effective in the presence of coordinated attacks. It would be a fascinating experiment for future work to test out the ensemble-based technique on more difficult attack models and other decentralized algorithms.